\documentclass[letterpaper]{article} 
\usepackage[preprint]{aaai2027}  
\usepackage[hyphens]{url}  
\usepackage{graphicx} 
\usepackage{natbib}  
\usepackage{caption} 
\usepackage{algorithm}
\usepackage{algorithmic}
\usepackage{amsmath}
\usepackage{amssymb}
\usepackage{multirow}

\usepackage{newfloat}
\usepackage{listings}
\DeclareCaptionStyle{ruled}{labelfont=normalfont,labelsep=colon,strut=off} 
\floatstyle{ruled}
\newfloat{listing}{tb}{lst}{}
\floatname{listing}{Listing}

\usepackage{booktabs}

\title{Interactive TTS: Dynamic Speaking Style Adaptation for Expressive Speech Synthesis}
\author{
    Wenjie Tian\textsuperscript{\rm 1},
    Kangxiang Xia\textsuperscript{\rm 1},
    Jingbin Hu\textsuperscript{\rm 1},
    Xinfa Zhu\textsuperscript{\rm 1},
    HangRui Hu\textsuperscript{\rm 1},
    Ziyue Jiang\textsuperscript{\rm 1},
    Kexin Huang\textsuperscript{\rm 1},
    Ting He\textsuperscript{\rm 1},
    Lei Xie\corresponding,
    Jin Xu\textsuperscript{\rm 1}\corresponding
}
\affiliations{
    \textsuperscript{\rm 1}\includegraphics[height=0.4cm]{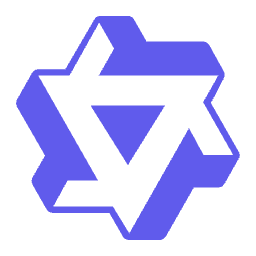} Alibaba ATH Token Foundry\\
}

\begin{document}

\maketitle

\begin{abstract}
Dynamic speaking style adaptation in multi-turn multimodal interaction remains a major challenge for text-to-speech (TTS) systems. Existing context-aware TTS (CTTS) methods typically map dialogue context to speech in an end-to-end manner. Such implicit modeling makes contextual style decisions difficult to supervise, while the entanglement of style, timbre, and content often leads to weak instruction-following and severe timbre drift across turns. 
To overcome these limitations, we propose Interactive TTS, a dynamic, style-adaptive framework for contextually appropriate and speaker-consistent speech generation. Interactive TTS decouples the process by explicitly modeling contextual style decisions as executable instructions. To bridge the gap between style decisions and speech generation, we introduce Iterative Rejection Sampling Fine-Tuning (Iterative RSFT) and Context-Aware Direct Preference Optimization (CADPO), which significantly enhance instruction-following and align the generated speech with conversational contexts. Extensive experiments demonstrate that Interactive TTS outperforms state-of-the-art models on VStyle and SpeechParaling-Bench. Demo is available at
\url{https://wjtian-wonderful.github.io/InteractiveTTS/}

\end{abstract}

\section{Introduction}
Recent advances in text-to-speech (TTS)~\cite{f5tts,qwen3tts,seedtts,indextts2,sparktts,llasatts,Cosyvoice2tts} have enabled highly natural speech synthesis for isolated utterances. 
However, speech generation in real-world interactive settings requires more than simply reading a sentence aloud. 
A practical system must dynamically adjust \emph{how} it speaks according to user intent, dialogue state, and environmental conditions, while maintaining a consistent speaker identity across turns. 
For example, it may need to sound more urgent in a dangerous situation or gentler when responding to a frustrated user. We refer to this capability as \emph{dynamic speaking style adaptation}, where the desired speaking style changes across dialogue turns rather than remaining fixed throughout an interaction.

\begin{figure}[t]
\centering
\includegraphics[width=\columnwidth]{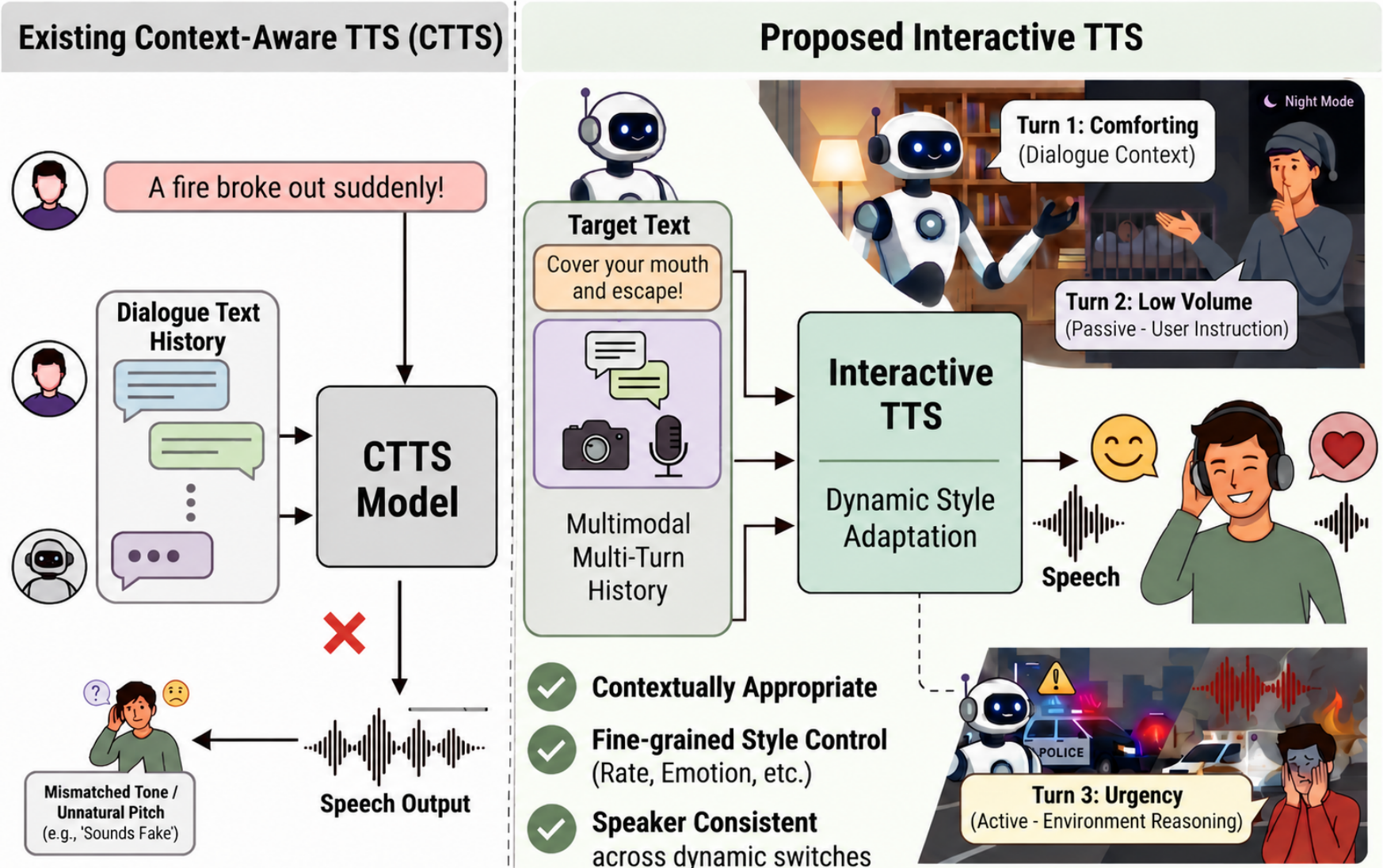}
\caption{Comparison between existing context-aware TTS (CTTS) and the proposed Interactive TTS. CTTS implicitly maps dialogue history to speech, often yielding mismatched expression. Interactive TTS dynamically adapts speaking style, while producing contextually appropriate speech.}
\label{fig:overall}
\end{figure}

Existing context-aware TTS (CTTS) systems~\cite{h4ctts,m2ctts,gpt_ctts,radkactts,DiffCSSctts,cong2021controllablectts,DailyTalk,gpttalker_ctts} take an important step toward this goal by incorporating dialogue history or multimodal context into speech generation, using conversational history as an auxiliary input. However, they typically rely on an implicit end-to-end mapping from context to speech. 

Despite their progress, existing CTTS systems still suffer from three critical bottlenecks that prevent them from being truly interactive:
First, in practice, style cues are often sparse and indirect in multimodal interaction. Sometimes the user explicitly specifies how the system should speak, e.g., ``speak more slowly.'' In many other cases, the desired style must be inferred from contextual signals such as linguistic semantics, acoustic prosody, or visual scenes. When these contextual style decisions are absorbed into a single implicit acoustic mapping, they become difficult to supervise, interpret, and control.
Second, training objectives mainly optimize alignment with the groundtruth speech, rather than whether the generated expression is contextually appropriate or instruction-following. As a result, the model may produce speech that sounds natural but expresses the wrong attitude, emotion, or urgency.
Third, speaking style is strongly entangled with linguistic content and speaker identity under end-to-end mapping paradigms.
Strong and frequent style changes may further interfere with speaker-related characteristics, resulting in speaker identity or timbre drift across turns. 
These limitations suggest that contextual style decision-making and acoustic realization should be modeled and aligned more explicitly.

To address these challenges, as illustrated in Fig.~\ref{fig:overall}, we propose \emph{Interactive TTS}: given the target text and interaction history, it generates context-appropriate and speaker-consistent speech while dynamically adapting style according to both explicit instructions and implicit multimodal context across dialogue turns.
Based on an instruction-based generator, we additionally introduce a Context-to-Instruction (C2I) module to explicitly model contextual style and produce an executable instruction describing how the target utterance should be spoken. Such an interface makes contextual style decisions explicit and enables direct supervision and evaluation of style inference independently of the downstream acoustic output.
We then enhance style inference through structured CoT reasoning from sparse multimodal information and On-Policy Self-Distillation (OPSD)~\cite{opsd}, which mitigates the degradation of multimodal reasoning capabilities.
We additionally post-train the instruction-conditioned generator along three dimensions: instruction faithfulness, speaker consistency, and contextual appropriateness. Specifically, Iterative Rejection Sampling Fine-Tuning (RSFT) uses evaluator-selected model rollouts to improve style instruction following while preserving speaker identity. Context-Aware Direct Preference Optimization (CADPO)~\cite{dpo} further incorporates the original interaction context into preference learning, encouraging speech that not only follows the style instruction but also fits the current conversational situation.

Our main contributions are summarized as follows:

\begin{itemize}
    \item We propose \emph{Interactive TTS}, which achieves dynamic style adaptation in multi-turn multimodal interactions while generating contextually appropriate and speaker-consistent speech.

    \item We propose a C2I perception framework that explicitly introduces contextual style decisions as executable instructions, with CoT supervision and OPSD improving multimodal style inference.
    
    \item We develop a multi-objective post-training approach to faithfully
    realize these style decisions in speech. Iterative RSFT improves instruction
    following while preserving speaker identity, and CADPO further aligns the
    generated speech with the original interaction context.
    
    \item Experiments on multiple perception and end-to-end benchmarks demonstrate superior performance over strong open-source and commercial baselines.
\end{itemize}

\section{Related Work}
\label{sec:related_work}

\subsection{Context-Aware and Conversational TTS}
To synthesize natural speech that conforms to conversational contexts, early work~\cite{DailyTalk} employs a pre-trained BERT to encode historical dialogue turns. 
Recently, generative and multimodal paradigms have further pushed the boundaries of this field.
M²-CTTS\cite{m2ctts} and H4C-TTS\cite{h4ctts} introduce multi-scale cross-modal modeling of dialogue text and historical speech.
DiffCSS \cite{DiffCSSctts} leverages diffusion probabilistic models to predict diverse prosody under long-context constraints, and RADKA-CSS \cite{radkactts} incorporates a retrieval-augmented mechanism to aggregate historical dialogue knowledge. 
GPT-talker~\cite{gpttalker_ctts}, CoT-TTS~\cite{cot_tts_ctts} and the talker of Qwen3omni~\cite{qwen3omni} introduce history context as condition in autoregressive generation.
Furthermore, HarnessTTS \cite{Harnesstts_ctts} utilizes external LLMs to synthesize contextual information. 
While some explicit control models, such as EmoOmni~\cite{emoomnitts}, attempt to address expressiveness, but they remain limited to single-turn scenarios and have control capabilities limited to the emotional dimension.

\subsection{Instruction-Conditioned TTS and End-to-End Spoken LLMs}
To enable fine-grained and explicit control over speech generation, instruction-conditioned TTS has received increasing attention. Representative systems such as Higgs Audio TTS\cite{higgsaudio2025}, VoxCPM2\cite{VoxCPM2}, and Qwen3-TTS~\cite{qwen3tts} use natural-language descriptions to control expressive attributes such as emotion, speaking rate, and speaking style. These studies demonstrate that language instructions provide a flexible and unified interface for expressive synthesis. Despite their powerful capabilities in style control, they perform unsatisfactorily in complex contexts due to the lack of a planner capable of integrating historical information.

End-to-end spoken LLMs~\cite{qwen2.5omni,qwen3omni,gpt4o,kimi-audio} further integrate multimodal understanding and speech synthesis into a unified framework, enabling natural spoken interactions. Nevertheless, their end-to-end formulation typically entangles contextual understanding, linguistic response generation, and acoustic realization, resulting in limited controllability and suboptimal style adaptation.

\subsection{Context-Aware Benchmarks and Datasets}

Evaluating the conversational appropriateness of speech in continuous interactions remains challenging. 
Most prior context-aware TTS studies rely heavily on subjective listening tests, which are costly and time-consuming. Recently, VStyle~\cite{vstyle} is a LLM-as-a-Judge benchmark for evaluating spoken dialogues, with an emphasis on acoustic style, role-playing, and implicit empathy. 
Similarly, SpeechParaling-Bench~\cite{SpeechParaling} provides a systematic pairwise evaluation protocol covering situational adaptation, dynamic paralinguistic control, and multi-turn variation.

On the data side, existing conversational speech datasets such as DailyTalk~\cite{DailyTalk} and NCSSD~\cite{gpttalker_ctts} are largely derived from recordings of everyday conversations or television and film content. Style variation in these corpora is dominated by mild prosodic entrainment, whereas expressive shifts triggered by situational changes are scarce, offering limited supervision for the fine-grained style control studied in this work. 
Instruction-following TTS corpora~\cite{instructts_data,instructts_data2} pair speech with explicit human-specified style descriptions but discarding the interaction history from which styles should be inferred. 

\section{Method}
\subsection{System Overview}

\begin{figure*}[t]
\centering
\includegraphics[width=\textwidth]{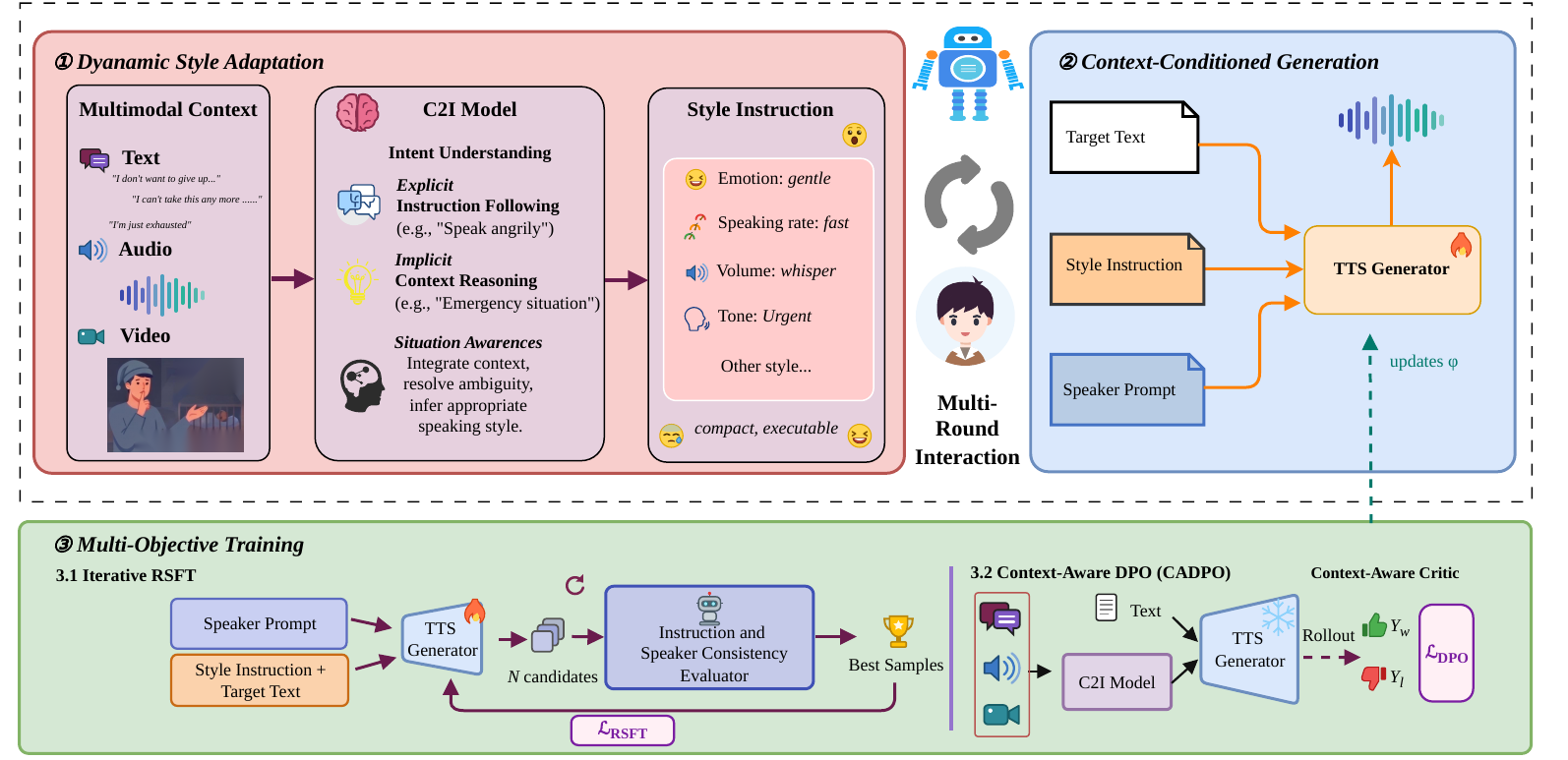}
\caption{Architecture of the proposed Interactive TTS framework. \textbf{1. Dynamic Style Adaptation:} The C2I model parses multimodal context into a compact, executable style instruction. \textbf{2. Context-Conditioned Generation:} The TTS generator synthesizes speech conditioned on the target text, style instruction, and a reference speaker prompt. \textbf{3. Multi-Objective Training:} The generator is aligned via Iterative RSFT for speaker consistency and instruction following, followed by CADPO to ensure contextual appropriateness.}
\vspace{-6pt}
\label{fig:model}
\end{figure*}

At dialogue turn $t$, Interactive TTS receives a target text $X_t$ and the interaction history

\begin{equation}
\mathcal{C}_t = \{M_1, M_2, \ldots, M_t\},
\end{equation}
where each $M_i$ may contain linguistic, acoustic, and visual observations. 
The context may contain an explicit style request, implicit evidence of an appropriate speaking style, or both.
Given a reference-speaker prompt $S_{\mathrm{ref}}$, the task is to generate speech $Y_t$ that is contextually appropriate while preserving the reference speaker's identity.

As shown in Fig.~\ref{fig:model}, we decompose the task into perception and expression. First, a Context-to-Instruction (C2I) model $P_\theta$ converts the interaction context and target text into a human-readable style instruction:
\begin{equation}
I_t \sim P_\theta(I_t \mid \mathcal{C}_t, X_t).
\label{eq:c2i}
\end{equation}
The instruction provides a unified representation for both user-specified and context-inferred style requirements. A speech generator $G_\phi$ subsequently realizes the target text according to the instruction and reference speaker:
\begin{equation}
Y_t \sim
G_\phi(Y_t \mid X_t, I_t, S_{\mathrm{ref}}).
\label{eq:generation}
\end{equation}
Here, $Y_t$ is represented as a sequence of discrete speech tokens.

\subsection{Context-to-Instruction}
Our C2I model is built on the thinker of Qwen-2.5omni-7B.
Style cues in multimodal interaction are often sparse, indirect, and
distributed across turns. Rather than directly predicting the final
instruction, C2I first generates a rationale $R_t$ that aggregates
style-relevant evidence and then produces an executable instruction
$I_t$.

Moreover, in practice, it may be insufficiently stable for a multimodal LLM to infer style directly from raw audio and video, because the model may miss or hallucinate style-relevant cues, and the automatically constructed data pipeline for the generated style may also introduce noise. Inspired by OPSD, we introduce additional audio-visual captions as supplementary information, providing clearer evidence for the same interaction and improving style prediction accuracy.

\subsubsection{Rationale and Instruction Supervision}

We represent $I_t$ as a compact natural-language description of the desired expression. It may specify emotion, emotional intensity, speaking rate, volume, tone/attitude, or other relevant paralinguistic attributes. For example:
\begin{quote}
\textit{``Speak slightly faster, at a moderately lower volume, with a reassuring tone.''}
\end{quote}
Unlike a fixed categorical style space, this representation can express combinations of attributes at different levels of granularity.

Due to the lack of such paired context-instruction data, we use Gemini-2.5-Pro~\cite{gemini2.5} to construct initial supervision data. 
Given $(\mathcal{C}_t,X_t)$, the external MLLM first generates a rationale $R_t$ that identifies and integrates style-relevant contextual evidence, and then summarizes the desired expression as $I_t$.
This produces training tuples
\begin{equation}
(\mathcal{C}_t, X_t, R_t, I_t).
\end{equation}
As illustrated in Stage 1 of Fig.~\ref{fig:model}, the rationale serves only as intermediate supervision for contextual analysis; it is neither treated as an independent style representation nor passed to the speech generator.

During training, we concatenate the rationale and instruction into the target sequence $A_t=[R_t;I_t]$ and minimize the following autoregressive negative log-likelihood:
\begin{equation}
\mathcal{L}_{\mathrm{SFT}}(\theta)
=
-
\sum_{j=1}^{|A_t|}
\log
P_\theta
\left(
a_{t,j}
\mid
a_{t,<j},\mathcal{C}_t,X_t
\right),
\end{equation}
\noindent
where $a_{t,j}$ denotes the $j$-th token of $A_t$. This objective directly supervises both evidence aggregation and final instruction generation.

\subsubsection{On-Policy Self-Distillation}

Let $T_{\mathrm{priv},t}$ denote the additional captions for the raw
multimodal context $\mathcal{C}_t$. We obtain $T_{\mathrm{priv},t}$
using Qwen-Omni-Plus, which generates audio captions describing
acoustic and paralinguistic attributes together with ASR transcripts,
as well as video captions summarizing the visual scene.

During training, the student conditions only on
$(\mathcal{C}_t,X_t)$, whereas the teacher additionally observes
$T_{\mathrm{priv},t}$. We first sample a rationale--instruction
sequence from the current student:
\begin{equation}
\widetilde{A}_t
\sim
p_{\mathrm{stu}}(\cdot\mid\mathcal{C}_t,X_t),
\qquad
\widetilde{A}_t=[\widetilde{R}_t;\widetilde{I}_t].
\end{equation}
We then minimize the token-level divergence between the teacher and
student distributions along the sampled trajectory:
\begin{equation}
\begin{aligned}
\mathcal{L}_{\mathrm{OPSD}}
=&\frac{1}{|\widetilde{A}_t|}
\sum_{j=1}^{|\widetilde{A}_t|}
\mathrm{KL}\Big(
\operatorname{sg}\big[
p_{\mathrm{tea}}(
\cdot\mid\widetilde{a}_{t,<j},
\mathcal{C}_t,X_t,T_{\mathrm{priv},t})
\big]
\\
&\qquad\qquad\quad \|
p_{\mathrm{stu}}(
\cdot\mid\widetilde{a}_{t,<j},
\mathcal{C}_t,X_t)
\Big),
\end{aligned}
\end{equation}
where $\operatorname{sg}[\cdot]$ denotes the stop-gradient operation.

At inference time, C2I generates $[R_t;I_t]$ from
$(\mathcal{C}_t,X_t)$ without additional captions, and only $I_t$ is
passed to the speech generator.

\subsection{Generation Framework}

We build on the pretrained Qwen3-TTS-CustomVoice~\cite{qwen3tts}, a TTS foundation model that supports style modification while preserving the speaker’s timbre.
Then we mainly perform three sequential alignment stages. 
Dynamic style changes across dialogue turns may cause speaker-identity
drift. At the very beginning, we place the reference speech at the system-prompt segment
to provide persistent speaker condition, 
Iterative Rejection Sampling Fine-Tuning (RSFT) then improves instruction execution while constraining speaker similarity. 
Context-Aware Direct Preference Optimization (CADPO) then distinguishes among instruction- and speaker-valid realizations according to the original interaction context.

\subsubsection{Persistent Reference-Speaker Conditioning}
We convert the fixed reference
utterance $S_{\mathrm{ref}}$ into acoustic tokens, insert it into the
system-prompt segment, and reuse it at each dialogue turn. 
We find that
this persistent conditioning effectively improves speaker consistency
during multi-turn synthesis. Since the pretrained generator was not
originally trained with this conditioning format, we briefly adapt it
using instruction-conditioned and same-speaker, different-utterance data
before RSFT.

\subsubsection{Iterative Rejection Sampling Fine-Tuning}
The instruction-conditioned generator may fail to follow challenging
style instructions, while stronger style variation can lead to
speaker-identity drift. We therefore apply Iterative Rejection Sampling
Fine-Tuning (RSFT), selecting samples that jointly achieve better
instruction faithfulness and speaker consistency.

Let $\phi^{(k)}$ denote the generator parameters at iteration $k$. For each 
training input consisting of text $X_t$, style instruction $I_t$, and reference 
speech $S_{\mathrm{ref}}$, we first sample $N$ candidate speech outputs from the 
current generator. 
To filter and select the best candidate, we employ an external multimodal judge 
(Gemini-2.5-Pro) to perform pairwise comparisons among the $N$ candidates. 
Given the style instruction and reference speech, the judge evaluates which 
candidate in each pair better follows the instruction while preserving the 
reference speaker's identity. We then conduct a round-robin tournament among 
all candidates and select the one with the highest number of pairwise wins. 
Gemini-2.5-Pro achieves approximately 78\% pairwise preference agreement with 
human evaluators, indicating that it provides a sufficiently reliable 
supervision signal for large-scale rollout filtering.
The selected best candidates from all inputs are collected to form the fine-tuning 
dataset $\mathcal{D}_{\mathrm{RSFT}}^{(k)}$. The generator is then fine-tuned by 
minimizing the standard negative log-likelihood loss:

\resizebox{\linewidth}{!}{$\displaystyle
\mathcal{L}_{\mathrm{RSFT}}(\phi)
=
-
\mathbb{E}_{(X_t,I_t,S_{\mathrm{ref}},\hat{Y}^*)\sim\mathcal{D}_{\mathrm{RSFT}}^{(k)}}
\left[
\log G_\phi(\hat{Y}^*\mid X_t,I_t,S_{\mathrm{ref}})
\right],
$}
\label{eq:rsft_loss}

\noindent
where $\hat{Y}^*$ represents the selected optimal candidate. 

The updated generator is then used to resample candidates for the next
iteration. We perform 5 rounds of RSFT, progressively improving
instruction faithfulness while limiting speaker-identity drift.

\subsubsection{Context-Aware DPO}

Instruction- and speaker-valid candidates may still differ in their appropriateness for the current interaction. 
We therefore introduce Context-Aware Direct Preference Optimization (CADPO) to distinguish among such realizations.

A context-aware critic then compares candidate pairs depending on the original interaction context $\mathcal{C}_t$, target text $X_t$, and instruction $I_t$. This produces preference pairs $(Y_t^w,Y_t^l)$, where $Y_t^w$ is judged to be more contextually appropriate than $Y_t^l$.
Let $G_{\phi_{\mathrm{ref}}}$ be a frozen copy of the RSFT-aligned generator before CADPO. We define
\begin{equation}
h_\phi(Y_t)
=
\beta
\log
\frac{
G_\phi(
Y_t
\mid
X_t,I_t,S_{\mathrm{ref}}
)
}{
G_{\phi_{\mathrm{ref}}}(
Y_t
\mid
X_t,I_t,S_{\mathrm{ref}}
)
},
\label{eq:cadpo_implicit_reward}
\end{equation}
where $\beta$ controls the preference optimization strength. The CADPO objective is
\begin{equation}
\mathcal{L}_{\mathrm{CADPO}}(\phi)
=
-
\mathbb{E}_{\mathcal{D}_{\mathrm{pref}}}
\left[
\log
\sigma
\left(
h_\phi(Y_t^w)-h_\phi(Y_t^l)
\right)
\right].
\label{eq:cadpo_loss}
\end{equation}

The interaction context is used only by the critic to construct preference pairs, while the generator remains conditioned on $(X_t,I_t,S_{\mathrm{ref}})$. CADPO therefore transfers context-aware supervision into the generator without requiring the multimodal context or critic at inference time.

\subsection{Data Construction and Training}

Overall, training proceeds sequentially. We first optimize C2I with rationale--instruction supervision and OPSD, then align the pretrained speech generator through iterative RSFT, followed by CADPO.
Most existing multimodal dialogue datasets~\cite{DailyTalk, gpttalker_ctts} are derived from television or film content and mainly contain everyday conversations, limiting their expressive diversity for the style control scenarios studied in this work. To address this issue, we construct a new multi-turn multimodal dialogue corpus with context-dependent expressive behaviors. The dataset includes 22K text-only turns, 47K audio turns (149 hours), and 12K video turns (32 hours), generated using Seedance2.0, Qwen3-TTS, and Gemini-3.1-Flash-TTS-Preview. All of these data are used to train the C2I module. Detailed annotation procedures and statistics are provided in Appendix~1.
For generator alignment, we build on Qwen3-TTS-12Hz-1.7B-CustomVoice. We construct approximately 2,000 hours of instruction-following synthesis data for RSFT by randomly sampling speakers from Emilia and synthesizing speech with diverse control instructions. For CADPO, we further reuse the C2I contextual data to generate TTS rollouts and filter them with a context-aware preference pipeline, resulting in approximately 200 hours of paired preference data.

At inference time, C2I predicts instruction from history context and target text, and only the instruction is passed to the generator, which synthesizes final speech conditioned on reference speaker, instruction and target text. 
The RSFT judge, the context-aware preference critic and the additional captions are required only during training.

\section{Experiments}

\begin{table*}[t]
\centering
\scriptsize
\setlength{\tabcolsep}{2pt}
\begin{tabular}{l|ccccccc c|ccc c|cc c|cccc c|c}
\toprule
\multirow{2}{*}{Model}
  & \multicolumn{8}{c|}{Acoustic Attributes $\uparrow$}
  & \multicolumn{4}{c|}{Instruction $\uparrow$}
  & \multicolumn{3}{c|}{Role-Play $\uparrow$}
  & \multicolumn{5}{c|}{Empathy $\uparrow$}
  & \multirow{2}{*}{Overall} \\
\cmidrule(lr){2-9} \cmidrule(lr){10-13} \cmidrule(lr){14-16} \cmidrule(lr){17-21}
  & Age & Speed & Gend. & Emot. & Pitch & Vol. & Comp. & \textbf{Avg}
  & Emot. & Style & Vari. & \textbf{Avg}
  & Scen. & Char. & \textbf{Avg}
  & Anger & Sad. & Anx. & Joy & \textbf{Avg}
  &  \\
\midrule
\multicolumn{22}{c}{\textbf{Context-Aware TTS Models}} \\
\midrule
\textbf{Interactive TTS}
& \textbf{3.98} & \textbf{4.30} & \underline{3.23} & \textbf{4.62} & 3.25 & \underline{4.33} & 3.22 & \underline{3.58}
& \textbf{4.33} & \textbf{4.42} & \textbf{3.88} & \textbf{4.19}
& \underline{3.63} & \underline{3.58} & \underline{3.61}
& \textbf{3.78} & \underline{3.83} & \underline{3.52} & 3.06 & \underline{3.55}
& \textbf{3.82} \\
Qwen3-Omni Talker
& 3.15 & 3.23 & 3.08 & 3.40 & 2.77 & 3.00 & 2.96 & 3.03
& 3.40 & 3.96 & 3.50 & 3.60
& 3.09 & 3.06 & 3.08
& 2.73 & 3.39 & 3.05 & 2.81 & 2.96
& 3.22 \\
HarnessTTS
& 3.71 & 3.88 & \textbf{3.35} & \underline{4.19} & \textbf{3.77} & \textbf{4.53} & \textbf{3.43} & \textbf{3.67}
& 4.02 & 3.90 & 3.18 & 3.67
& 3.59 & 3.24 & 3.43
& 3.01 & 3.29 & 2.56 & 2.26 & 2.78
& 3.47 \\
CoT-TTS
  & 2.93 & 3.05 & 2.96 & 3.26 & 2.57 & 2.95 & 2.74 & 2.84
  & 3.31 & 3.24 & 2.84 & 3.13
  & 2.86 & 2.80 & 2.83
  & 2.20 & 2.33 & 1.82 & 1.84 & 2.05
  & 2.71 \\
GPT-Talker
& 2.83 & 2.85 & 2.77 & 2.96 & 2.61 & 3.10 & 2.64 & 2.75
& 3.23 & 3.11 & 2.88 & 3.07
& 2.16 & 2.23 & 2.19
& 2.23 & 2.57 & 1.87 & 1.94 & 2.15
& 2.54 \\
\midrule
\multicolumn{22}{c}{\textbf{End-to-End Spoken LLMs}} \\
\midrule
Qwen-Audio-3.0-Realtime~\footnote{\url{https://www.alibabacloud.com/help/en/model-studio/qwen3-5-livetranslate-flash-realtime}}
& 3.46 & \underline{4.08} & 3.12 & 3.85 & 3.23 & 3.92 & 3.30 & 3.46
& \underline{4.06} & \underline{4.14} & \underline{3.73} & \underline{3.96}
& \textbf{3.71} & \textbf{3.73} & \textbf{3.72}
& 3.06 & 3.15 & 3.18 & 3.07 & 3.11
& \underline{3.64} \\
GPT-4o Audio
& 3.35 & 3.45 & 3.15 & 3.62 & 3.25 & 4.05 & \underline{3.32} & 3.40
& 3.94 & 3.93 & 3.57 & 3.80
& 2.97 & 3.34 & 3.14
& \underline{3.60} & \textbf{3.92} & \textbf{3.67} & \underline{3.59} & \textbf{3.68}
& 3.53 \\
Doubao-S2S
& \underline{3.88} & 3.75 & 3.10 & 3.87 & \underline{3.35} & 3.95 & \underline{3.32} & 3.48
& 4.00 & 4.07 & 3.24 & 3.74
& 2.99 & 2.94 & 2.97
& 3.51 & 3.44 & 3.20 & \textbf{3.67} & 3.47
& 3.46 \\
GPT-4o-mini Audio
& 2.85 & 3.77 & 3.00 & 3.83 & 2.85 & 3.67 & 2.96 & 3.14
& 3.69 & 3.82 & 3.33 & 3.59
& 2.62 & 3.01 & 2.80
& 3.12 & 3.31 & 2.84 & 3.43 & 3.18
& 3.23 \\
Qwen3.5-Omni-Plus
& 2.77 & 2.88 & 2.67 & 3.48 & 2.30 & 2.95 & 2.56 & 2.70
& 2.94 & 3.03 & 2.60 & 2.84
& 2.49 & 2.69 & 2.58
& 3.28 & 3.76 & 3.25 & 3.22 & 3.36
& 2.85 \\
Kimi-Audio
& 3.10 & 2.75 & 2.58 & 3.46 & 2.12 & 3.23 & 2.48 & 2.67
& 2.64 & 2.85 & 2.55 & 2.67
& 2.22 & 1.77 & 2.02
& 2.98 & 2.86 & 2.31 & 3.20 & 2.86
& 2.57 \\
Step-Audio
& 2.71 & 2.48 & 2.23 & 3.04 & 2.20 & 2.80 & 2.41 & 2.49
& 2.62 & 2.97 & 2.43 & 2.65
& 2.16 & 2.19 & 2.17
& 2.53 & 3.10 & 2.33 & 2.81 & 2.68
& 2.51 \\
Qwen2.5-Omni
& 2.88 & 2.33 & 2.25 & 3.00 & 2.15 & 2.25 & 2.38 & 2.42
& 2.87 & 2.66 & 2.38 & 2.63
& 1.88 & 1.84 & 1.86
& 2.88 & 2.80 & 3.05 & 3.37 & 3.02
& 2.48 \\
Qwen3-Omni
& 2.71 & 2.31 & 2.08 & 2.92 & 2.42 & 2.02 & 1.93 & 2.17
& 1.90 & 2.13 & 1.99 & 2.01
& 2.66 & 2.79 & 2.72
& 2.90 & 3.44 & 2.77 & 2.94 & 3.00
& 2.42 \\
Baichuan-Audio
& 2.56 & 2.38 & 3.10 & 2.50 & 2.05 & 2.45 & 2.61 & 2.56
& 2.22 & 2.39 & 1.89 & 2.15
& 2.01 & 1.95 & 1.98
& 1.88 & 2.58 & 1.69 & 2.60 & 2.17
& 2.21 \\
\bottomrule
\end{tabular}
\caption{Comparison on VStyle benchmark. \textbf{Overall} is averaged over all evaluation samples. The \textbf{Acoustic Attributes} average follows the official VStyle weighting, where the Composite dimension contributes one half of the category score and the other six dimensions share the remaining half. The \textbf{Instruction}, \textbf{Role-Play}, and \textbf{Empathy} averages are the unweighted means of their respective sub-dimensions. Our system uses C2I to infer an explicit style instruction from the benchmark input and then synthesize speech with an instruction-conditioned generator.}
\vspace{-12pt}
\label{tab:vstyle-finegrained}
\end{table*}

\subsection{Experimental Setup}

\subsubsection{Implementation details.}


All models are trained on 32 NVIDIA A100 GPUs with 80 GB memory each. 
For C2I, the supervised fine-tuning stage was conducted for 3 epochs, followed by 1 epoch of OPSD training. The context length is set to 32K for every training step. 
RSFT is trained for 5 rounds. 
In addition, CADPO is performed for 2 epochs.
The context length is set to 8K for every training step.

\subsubsection{Benchmarks}
Since VStyle and
SpeechParaling-Bench are originally designed for end-to-end
spoken LLM systems, we use Qwen-Audio-3.0-Realtime~\footnote{\url{https://www.alibabacloud.com/help/en/model-studio/qwen3-5-livetranslate-flash-realtime}} to generate
response texts from the input speech and conversational context. 
The same generated texts are then provided to all evaluated TTS systems to
ensure a fair comparison.

\textbf{VStyle.}
VStyle evaluates interactive speaking style in terms of context
adaptation, role-playing, and empathetic expression in open-domain
scenarios. A unified Gemini-2.5-Pro judge assigns scores from 1 to 5 and each score is averaged over the combined English and Chinese sets.
Additionally, we randomly sampled 2 items from each of the Chinese and English categories. All samples are evaluated by 17 human based on the vstyle scoring criteria. For further details, please refer to Appendix 5.

\textbf{SpeechParaling-Bench.}
SpeechParaling-Bench evaluates context-adaptive paralinguistic
generation in both English and Chinese. It contains three task types:
dynamic variation, situational adaptation, and paralinguistic control.
Following the official pairwise protocol, each candidate audio is compared with a fixed anchor audio. The two audios are randomly assigned to T1/T2, and a LALM judge assigns 0--3 scores and determines the winner for each dimension. The official anchors are Gemini for English and Doubao for Chinese, and the judge is Gemini-3.1-Pro-Preview.

\textbf{Multi-Turn Video Dialogue Evaluation.}
As a supplement to the VStyle and SpeechParaling-Bench, it lacks multimodal and multi-round testing. We evaluate the perception module independently on the multi-turn C2I-Video testset. 
It comprises 30 multimodal dialogues across six interaction scenarios, with 88 conversational turns in total. Among them, 50 key style-anchor turns require the system
to dynamically adapt its speaking style according to the multimodal context. We use Gemini-2.5-Pro to determine whether the style instruction predicted by the context-to-instruction (C2I) module is
consistent with the ground-truth instruction at each anchor turn, and
report the resulting accuracy. Further details are provided in
Appendix 2.

\textbf{Speech Generation Evaluation.}
We evaluate instruction faithfulness on an internal test set of 6,300
samples covering 21 speaking styles. The test set is evenly divided
into style-biased and neutral texts, where the latter requires the
target style to be conveyed primarily through the instruction rather
than textual semantics. Following
InstructTTSeval~\cite{InstructTTSEval} and VStyle, we use
Gemini-2.5-Pro as an automatic judge and report success rates for both
text conditions and their overall average.

We additionally evaluate speaker preservation using WavLM-based speaker
similarity (SIM)~\cite{chen2022wavlm}. Because speaking style and perceived timbre are
partially entangled, embedding-based similarity can be affected by
substantial style changes. We therefore complement SIM with a human
evaluation of speaker consistency on a category-balanced subset. For
each system, we select 2 samples from each of the 21 style categories,
one with neutral text and one with style-biased text. All samples are evaluated by 17 human raters.
Speech intelligibility is evaluated using WER/CER, following
SeedTTS~\cite{seedtts}. Further details are provided in
Appendix 3.

\subsubsection{Baselines}

We compare our system with several categories of strong baselines.

First, we include several open-sourced context-aware TTS models, including HarnessTTS~\cite{Harnesstts_ctts}, CoT-TTS~\cite{cot_tts_ctts}, and GPT-Talker~\cite{gpttalker_ctts}. 
As HarnessTTS only supports text-based context, we use the transcripts of the audio inputs as prompts.
In addition, we adapt the talker module of Qwen3-Omni into a context-aware TTS system by directly prefilling the text to be synthesized and feeding the history context into the thinker.

Second, since VStyle and SpeechParaling-Bench are proposed to evaluate end-to-end spoken LLMs, we also include several representative spoken LLM systems for performance reference, including Qwen3-Omni~\cite{qwen3omni}, Step-Audio~\cite{stepaudio2}, Kimi-Audio~\cite{kimi-audio}, the commercial streaming systems GPT-4o Audio~\cite{gpt4o}, GPT-4o-mini Audio~\cite{gpt4o}, Qwen3.5-Omni-Plus, and Doubao-S2S.

For the accuracy of the context-to-instruction (C2I) understanding component, we evaluate several strong commercial models, including Gemini-3.1-Pro-Preview~\cite{gemini2.5}, Doubao-Seed-2.0-Pro, Qwen3.5-Omni-Plus, and Gemini-3.5-Flash~\cite{gemini2.5}.

Finally, for controllable TTS generation models, we evaluate VoxCPM2~\cite{VoxCPM2}, Higgs Audio TTS~\cite{higgsaudio2025}, as well as our base model Qwen3TTS-CustomVoice~\cite{qwen3tts}.

\subsection{End-to-End Interactive Evaluation}

We first conduct end-to-end evaluation on VStyle and SpeechParaling-Bench. In this setting, the model must infer an appropriate speaking style from the interaction context and synthesize the target speech accordingly.

\subsubsection{VStyle Results}

Table~\ref{tab:vstyle-finegrained} reports the VStyle comparison under the official evaluation protocol. 
For the TTS baselines, subjective scores achieve a pearson correlation coefficient of 0.73 with the table's results, indicating the benchmark provides a useful reference.
Our model achieves the best overall performance among all TTS baselines. 
It also delivers competitive results on the \textit{Acoustic Attributes} and \textit{Instruction} groups, demonstrating its ability to faithfully follow explicit user controls.
Notably, Harness TTS achieves slightly superior control over pitch and gender because the retrieval-based method naturally preserves basic acoustic attributes more reliably than generative methods.
Moreover, in interactive settings, our model effectively leverages contextual information and achieves a state-of-the-art \textit{Empathy} score of 3.55 among existing TTS systems, reflecting its capability for implicit, context-aware adaptation.
Overall, the average scores for explicit control and \textit{Empathy} demonstrate that our model can effectively perform dynamic style control.

\begin{table}[t]
\centering
\setlength{\tabcolsep}{3pt}
\resizebox{\columnwidth}{!}{
\begin{tabular}{lcccc} 
\toprule
System & DynVar $\uparrow$ & SitAda $\uparrow$ & ParaCtrl $\uparrow$ & Overall $\uparrow$\\
\midrule
\multicolumn{5}{c}{\textbf{Context-Aware TTS Models}} \\
\midrule
\textbf{Interactive TTS} & 34.5 & \textbf{80.3} & \textbf{44.0} & \textbf{49.8} \\
Qwen3-Omni Talker & 16.0 & 63.7 & 11.0 & 21.7 \\
Harness TTS & 20.2 & 66.8 & 19.4 & 28.8 \\
CoT-TTS & 8.0 & 52.0 & 10.7 & 18.1 \\
GPT-Talker  & 10.3           & 47.4 & 11.1 & 17.9 \\
\midrule
\multicolumn{5}{c}{\textbf{End-to-End Spoken LLMs}} \\
\midrule
Qwen-Audio-3.0-Realtime & \textbf{36.6} & 62.9 & 31.2 & 37.9 \\
Qwen3.5-Omni-Plus & 6.7 & 39.2 & 7.2 & 13.3 \\
Qwen3-Omni & 1.3 & 47.4 & 1.3 & 10.1 \\
\bottomrule
\end{tabular}
}
\caption{Comparison on SpeechParaling-Bench. Values denote win rates (\%) against the official anchors for dynamic variation (DynVar), situational adaptation (SitAda), paralinguistic control (ParaCtrl), and Average (overall).}
\label{tab:speechparaling}
\end{table}

\subsubsection{SpeechParaling-Bench Results}
As shown in Table~\ref{tab:speechparaling}, Interactive TTS achieves the highest win rate of 80.3\% against the official anchors on the situational adaptation (SitAda) dimension, validating the accuracy of our dynamic style prediction. Notably, our training data contains no samples explicitly designed for dynamic variation or paralinguistic control. The strong performance on these dimensions is therefore mainly attributable to the generalization ability of the C2I module and the inherent capabilities of the base TTS generator, demonstrating the robustness of our framework beyond its training distribution.

\subsection{Perception Analysis}

\subsubsection{C2I-Video Evaluation}

\begin{table}[!t]
\centering
\footnotesize
\begin{tabular}{lc}
\toprule
System & Style Accuracy $\uparrow$ \\
\midrule
Gemini-3.1-Pro-Preview & 34/50 (68\%) \\
Doubao-Seed-2.0-Pro & 27/50 (54\%) \\
\textbf{Interactive TTS (C2I)} & 26/50 (52\%) \\
Qwen3.5-Omni-Plus & 23/50 (46\%) \\
Gemini-3.5-Flash & 21/50 (42\%) \\
Qwen3-Omni & 20/50 (40\%) \\
Qwen2.5-Omni-7B & 16/50 (32\%) \\
\bottomrule
\end{tabular}
\caption{Overall style accuracy on the C2I-Video testset. Results are reported as the number and percentage of correctly identified style anchors.}
\label{tab:seedance-anchor}
\end{table}


Table~\ref{tab:seedance-anchor} evaluates the style prediction capability of the C2I model given the interaction context. Our 7B C2I model performs comparably to the proprietary commercial model Doubao, highlighting the effectiveness of C2I as a specialized model for contextual style inference.

\subsection{Expression Alignment Analysis}

We compare our system with strong instruction-based TTS systems on the internal instruction-following test set.
As shown in Table~\ref{tab:rsft-flywheel}, our expression alignment raises the instruction success rate of the base Qwen3TTS from 79.5\% to 89.6\%, clearly surpassing the strong instruction-conditioned baselines VoxCPM2 (80.3\%) and Higgs Audio TTS (81.4\%).
Importantly, the improved style controllability does not come at the cost of speaker identity: Interactive TTS also achieves the highest Human-SIM (4.51) and SIM (0.694), while keeping WER comparable to the base model.
In contrast, Higgs Audio TTS exhibits a clear drop in speaker preservation (SIM 0.599, Human-SIM 3.86). These results confirm that RSFT yields stable gains on both objectives.

\begin{table}[!t]
\centering
\footnotesize
\setlength{\tabcolsep}{2pt}
\resizebox{\columnwidth}{!}{%
\begin{tabular}{l c c c c}
\toprule
Model / Setting & Inst. Success $\uparrow$ & Human-SIM $\uparrow$       & SIM $\uparrow$    & WER $\downarrow$   \\
\midrule
Qwen3TTS        & 79.5\%                   & $4.12 \pm 0.11$            & \underline{0.686} & 1.399\%            \\
Interactive TTS & \textbf{89.6\%}          & $\mathbf{4.51 \pm 0.08}$    & \textbf{0.694}    & \underline{1.360\%} \\
VoxCPM2         & 80.3\%                   & $\underline{4.37 \pm 0.09}$ & 0.681             & 1.425\%            \\
Higgs Audio TTS & \underline{81.4\%}       & $3.86 \pm 0.12$            & 0.599             & \textbf{1.201\%}    \\
\bottomrule
\end{tabular}%
}
\caption{Subjective and objective evaluation on speech generator. Instruction-following (Inst. Success), speaker-preservation (Human-SIM), and intelligibility(WER) results on our internal evaluation set.}
\label{tab:rsft-flywheel}
\end{table}

\subsection{Ablation Study}

\begin{table}[t]
\centering
\footnotesize
\setlength{\tabcolsep}{2pt} 
\begin{tabular}{l c c c}
\toprule
Variant & VStyle $\uparrow$ & C2I-Video $\uparrow$ & Human-SIM $\uparrow$ \\
\midrule
\textbf{Interactive TTS}    & \textbf{3.82} & \textbf{52.0\%}         & \textbf{4.51}        \\
\quad -- w/o CoT      & 3.77           & 48.0\%                  & --                   \\
\quad -- w/o OPSD           & 3.73           & 44.0\%                      & --                   \\
\quad -- w/o Response       & 3.79           & 46.0\%                  & --                   \\
\quad -- w/o CADPO          & 3.81            & --                  & 4.48                 \\
\quad \quad -- w/o RSFT           & 3.66            & --                  & 4.15                 \\
\midrule
Qwen3TTS    & 3.57           & --                      & 4.12                   \\
\bottomrule
\end{tabular}
\caption{Ablation of the proposed pipeline. Each row removes one component and is measured on its target metric.}
\vspace{-6pt}
\label{tab:ablation}
\end{table}

As shown in Table~\ref{tab:ablation}, all components contribute to the
final performance.
On the perception side, removing OPSD causes the largest drop in C2I
style accuracy on C2I-Video testset (52.0\% to 44.0\%), confirming that
caption-based OPSD is the key to reliable style
inference from raw multimodal context. 
Removing response content during C2I inference ('w/o Response') similarly degrades performance, decreasing C2I-Video to 46.0\%. 
We assume that the response provides additional conversational context, which helps the system produce speech that is not only stylistically appropriate but also interaction-aware.
For speaker consistency, removing CADPO slightly reduces Human-SIM from 4.51 to 4.48, while additionally removing RSFT causes a larger drop to 4.15. This shows that RSFT forms the basis of multi-turn speaker consistency, and CADPO further improves human-perceived identity stability, without noticeably compromising the SIM and C2I performance.

Finally, we evaluate Qwen3TTS on VStyle by replacing the input query with text.
Compared with the baseline Qwen3TTS, the full Interactive TTS system achieves clear gains on both overall style quality (3.82 vs. 3.57) and multi-turn speaker consistency (4.51 vs. 4.12). These improvements confirm the effectiveness of the proposed pipeline in jointly enhancing expressive style control and dialogue-level speaker stability.
\vspace{-8pt}

\section{Conclusion}

We introduced \emph{Interactive TTS}, a dynamic style-adaptive framework designed for multi-turn multimodal interaction. The key idea is to make contextual style decisions explicit: our framework first infers executable style instructions via C2I, and then realizes them through an instruction-conditioned generator aligned by RSFT and CADPO. 
This design improves controllability in contextual style adaptation, while reducing speaker drift under frequent style changes.
Results demonstrate that Interactive TTS produces more contextually appropriate and speaker-consistent speech than existing context-aware TTS and end-to-end spoken LLM baselines. 
A remaining challenge is to model richer, fine-grained style dynamics, especially variations within a single utterance and non-verbal vocal expressions. 
Furthermore, integrating these expressive capabilities into end-to-end spoken language models remains an important direction.

\bigskip
\newpage

\bibliography{aaai2027}


\end{document}